\documentclass[a4paper,11pt]{article}
\usepackage{jinstpub}
\usepackage{graphicx}
\usepackage{amsmath}

\usepackage{amssymb}
\usepackage{xcolor}

\title{Charged-particle topology reconstruction with an in-liquid SiPM array}

\author{H.~Kimku,}
\author{J.~S.~Chung,}
\author[1]{C.~Ha\note{Corresponding authors.},}
\author{T.~Z.~Huang,}
\author{J.~Kim,}
\author{S. A. ~Kim,}
\author{B.~C.~Koh,}
\author[1]{M.~S. ~Kwak,}
\author{S.~Lee,}
\author{Y.~J.~Lee}
\author{and J.~Seo}

\affiliation[]{Department of Physics, Chung-Ang University,\\
Seoul, 06974, Republic of Korea}

\emailAdd{chha@cau.ac.kr}
\emailAdd{maybe.minsu.kwak@gmail.com}

\abstract{Liquid scintillator detectors instrumented with photosensors inside the scintillation volume preserve local optical information that is largely lost in conventional boundary-readout geometries.
We demonstrate that this information is sufficient for charged-particle topology reconstruction using a sparse three-dimensional lattice of silicon photomultipliers.
After validating the Geant4 detector response against measured photon-count distributions, a simulation-trained, time-informed convolutional neural network reconstructs the entry and exit points of through-going muons with median residuals of 1.91~cm and 2.39~cm, respectively. The reconstructed endpoints are geometrically consistent with acceptance regions defined by external trigger counters in cosmic-ray muon data.
The same framework also reconstructs the production vertices of simulated positron starting-track events with a median residual of about 4.5~cm. These results establish the feasibility of topology-sensitive reconstruction using sparse in-liquid photosensor arrays in homogeneous liquid scintillator detectors.}


\begin{document}

\maketitle
\flushbottom

\section{Introduction}

Liquid scintillator detectors, including KamLAND, Borexino, SNO+, JUNO, and NEOS, are widely used in neutrino and rare-event experiments because they provide large target masses,
high light yields, and stable calorimetric responses~\cite{KamLAND:2004overview,Borexino:2008detector,SNOplus:2021experiment,JUNO:2021vlw,NEOS:2016}.
In conventional homogeneous detectors, photosensors are placed at the detector boundary. Multiple reflections, absorption, and long optical paths therefore reduce the local charge and timing information associated with charged-particle tracks before the photons are recorded~\cite{knoll,LiquidO:2021}.

We investigate whether a sparse photosensor lattice inside the scintillator retains enough local optical information to reconstruct charged-particle tracks.
The study examines the simulated detector response, reconstruction performance relative to a charge-only template baseline, and consistency with externally triggered cosmic-ray muon data.

Recent developments in silicon photomultiplier (SiPM) technology make it possible to place compact photosensors directly inside the scintillator volume~\cite{Renker:2006SiPM}.
Related topology-sensitive scintillator readout concepts have also been explored in other optical geometries~\cite{LiquidO:2021}.
This work uses an in-liquid photosensor detector as a test bed for reconstruction and includes only the geometry and response information required for that purpose.

We first validate the Geant4 photon-count response against cosmic-ray muon data. We then compare the SiPM-$r$ template method with charge-only and time-informed convolutional neural networks (CNNs) using simulated samples with truth information. In the SiPM-$r$ method, $r$ denotes the shortest distance between a candidate track and each SiPM.
The time-informed CNN is also tested with externally triggered cosmic-ray muon data to check real-data consistency.
In addition to through-going tracks, positron starting-track events are studied to test whether the same readout can support production-vertex reconstruction for a more localized event topology.

\section{CANDY detector}

The detector information needed for reconstruction and validation is summarized here.
The Chung-Ang University Neutrino Detection Yolk (CANDY) detector is a cubic liquid-scintillator detector with dimensions of $50\times 50\times 50~\mathrm{cm}^{3}$.
The detector vessel is constructed from 1-cm-thick acrylic panels and contains five internal acrylic support plates separated by 10~cm along the vertical direction.
Each plate holds 25 SiPMs arranged in a $5\times 5$ grid with 10~cm center-to-center spacing in both horizontal directions, giving 125 SiPM channels in a $5\times5\times5$ lattice.
All 125 SiPMs have their active faces pointing vertically upward, with their active surfaces parallel to the horizontal acrylic support plates.

\begin{figure}[htbp]
	\begin{center}
		\includegraphics[width=0.70\textwidth]{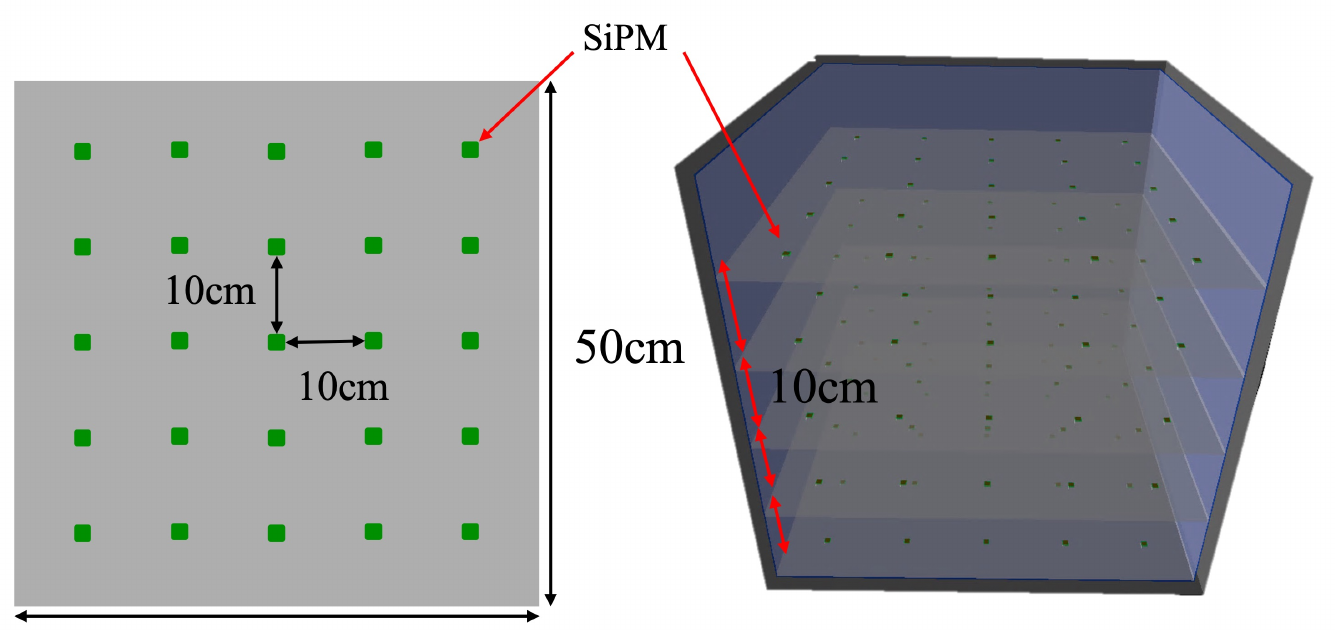}
	\end{center}
	\caption{
	  Schematic of the CANDY detector. SiPMs are mounted on transparent acrylic plates (left),
	  and the plates are installed with regular vertical spacing in the scintillator volume (right).
	}
\label{fig:detector}
\end{figure}

The detector was filled with a linear-alkylbenzene (LAB)-based liquid scintillator.
Internal detector surfaces were lined with enhanced specular reflector (ESR) film to improve optical collection efficiency and reduce position-dependent light loss~\cite{tyvek}.
Figure~\ref{fig:detector} shows the geometry and in-liquid SiPM lattice used in the reconstruction study.

Signals from the SiPMs~\cite{sipm} were digitized using custom fast analog-to-digital converter (fADC) modules operating at 62.5~MSa/s, corresponding to 16~ns per sample.
A channel was considered hit when its 128~ns sliding integral exceeded a fixed threshold, and an event trigger required at least five channels above threshold within a 1024~ns coincidence window.
For each trigger, 4096~ns of waveform data were recorded. The data used in this analysis were recorded at $-30^{\circ}\mathrm{C}$ to reduce dark-count noise, while baseline stability was monitored using periodic triggers.
Further details of the detector operation and temperature-dependent response are given in Ref.~\cite{CANDYLS}.

\section{Simulation and detector-response validation}

\subsection{Detector simulations}

The simulated cosmic-ray muon sample was produced using the Cosmic-Ray Shower Library (CRY)~\cite{Hagmann:2007CRY}, which provides atmospheric secondary-particle distributions for detector transport simulations.
Initial muon positions were sampled on a $3\times3~\mathrm{m}^2$ surface above the detector, and the particles were
propagated through the detector geometry to produce both through-going and stopping-muon events.

The detector response was simulated using Geant4~\cite{GEANT4:2002zbu,GEANT4:2016lxb}.
The simulation includes scintillation photon production, optical transport in the liquid scintillator,
		reflections from ESR reflector films, optical absorption, and photon detection by the in-liquid SiPM array.
The optical parameters adopted in the simulation are summarized in Table~\ref{tab:mcparams}.

For each SiPM, the simulation records the detected photon count and the mean arrival time of the detected photons.
The photon counts serve as inputs to both reconstruction methods, while the mean arrival times provide timing information for the time-informed CNN.
These samples were used to develop and evaluate the SiPM-$r$ template and CNN reconstruction methods presented in this work.

\begin{table}[htbp]
		\caption{Main optical and detector-response parameters used in the Geant4 simulated samples.}
	\label{tab:mcparams}
	\centering
	\smallskip
	\begin{tabular}{ll}
			Parameter             & Value                                \\
			\hline
			Scintillation yield   & $3000~\mathrm{photons/MeV}$          \\
			LAB absorption length & $10~\mathrm{m}$                      \\
			ESR reflectivity      & 0.85                                 \\
			Active volume         & $50\times50\times50~\mathrm{cm}^{3}$ \\
			SiPM lattice          & $5\times5\times5$ channels           \\
			Channel spacing       & $10~\mathrm{cm}$                     \\
		\end{tabular}
\end{table}

\subsection{Photon-count response validation}

\begin{figure}[tbp]
	\begin{center}
		\includegraphics[width=0.70\textwidth]{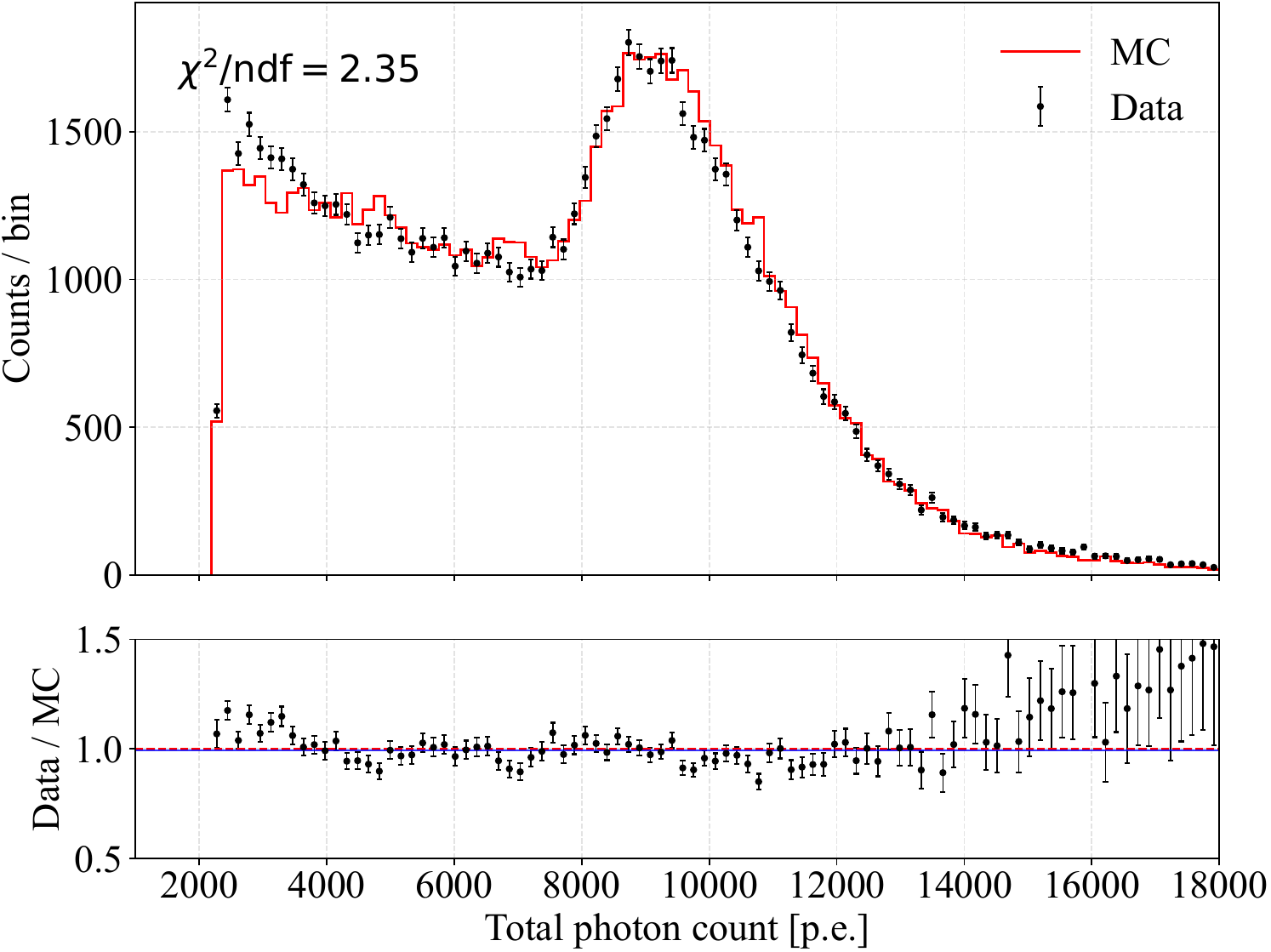}
	\end{center}
	\caption{
			Comparison of the measured total photon-count spectrum with the Geant4-based Monte Carlo (MC) prediction.
		The lower panel shows the data-to-MC ratio after the simulated photon-count scale correction and yield normalization.
	}
	\label{fig:datamc}
\end{figure}

The total photon-count response was validated against cosmic-ray muon data.
Figure~\ref{fig:datamc} compares the measured total photon-count distribution with the simulated response.

The simulation reproduces the dominant measured total-light scale after applying a single global correction factor of 2.2 to the simulated photon-count scale.
This correction is treated as an effective light-yield alignment rather than as a measurement of a single detector parameter.  It absorbs the combined uncertainties in SiPM photon-detection efficiency, scintillation yield, optical attenuation and reflection properties, and electronics calibration.
After applying this correction, the distribution-level agreement for the scaled spectra is quantified by a shape-only $\chi^{2}/\mathrm{ndf}$ of 2.35.
The shape-only $\chi^{2}$ is computed after normalizing the total Monte Carlo yield to the data yield, with a variance that includes both data and finite Monte Carlo statistics.

The comparison is most relevant in the photon-count region that dominates the triggered cosmic-ray sample used for reconstruction.
Deviations below the main comparison range are largely due to residual environmental gamma-ray backgrounds compounded by trigger-threshold effects.
At large total photon counts, the agreement is limited primarily by the finite Monte Carlo exposure, equivalent to 3600~s of detector live time,
and by the omission of high-multiplicity cosmic-ray muon-bundle events and uncorrelated accidental events from the simulated sample. These processes contribute to the highest-photon-count region in the experimental data but are not sufficiently represented in the current simulation.

Although the comparison does not constitute a complete absolute detector calibration,
it verifies the dominant photon-count response needed for the reconstruction study.
The subsequent reconstruction test using simulated events with truth information and the external-trigger validation provide complementary checks that the simulation-trained model gives physically consistent endpoint estimates.

\begin{figure}[htbp]
	\begin{center}
		\includegraphics[width=0.70\textwidth]{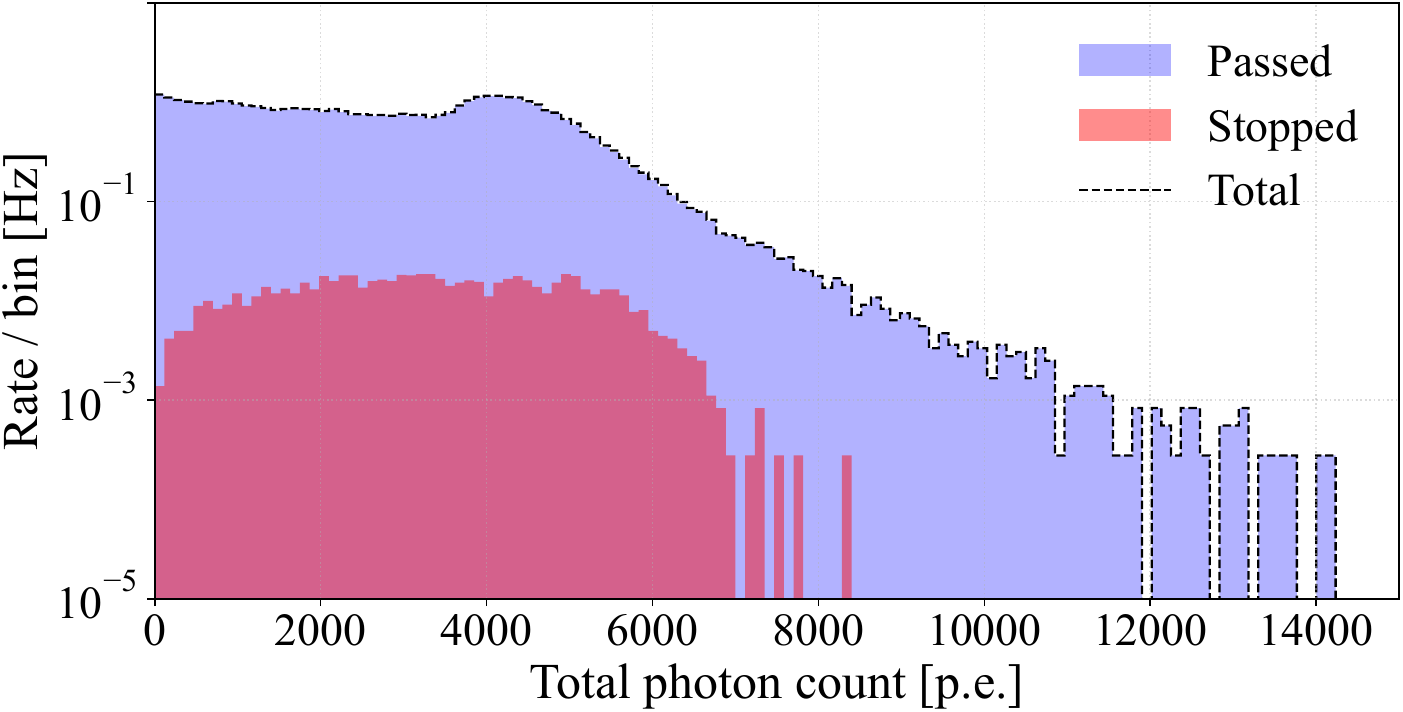}
	\end{center}
	\caption{
		Total photon-count distributions for CRY-generated cosmic-ray muon events.
		Truth information is used to distinguish through-going muons from stopping-muon candidates and to provide a simulation-based reference for stopping-muon-enriched sample selection.
		Stopping-muon candidates account for 1.55\% of triggered events in the simulated sample.
	}
	\label{fig:mcselection}
\end{figure}

A CRY muon sample with truth information was also used to study the total photon-count separation between through-going and stopping-muon candidates.
Figure~\ref{fig:mcselection} shows that stopping-muon candidates populate a lower-total-light region than through-going muons.
This comparison provides a simulation-based reference for stopping-muon selection in a separate timing study.

\section{Reconstruction methods}

\subsection{SiPM-$r$ template reconstruction}

The SiPM-$r$ template baseline uses Monte Carlo (MC) response templates to relate the photon count observed by each SiPM to its expected distance from a top-to-bottom track hypothesis. The best-fit track parameters $\hat{\theta}$ are obtained by combining the channel constraints,

\begin{equation}
	\hat{\theta}=\operatorname*{arg\,min}_{\theta}
	\sum_c \mathcal{C}_c\!\left[r_c(\theta);n_c\right],
\end{equation}

where $c$ indexes the SiPM channels, $r_c(\theta)$ is the shortest perpendicular distance between the position of SiPM $c$ and the track defined by hypothesis $\theta$, $n_c$ is the observed photon count, and $\mathcal{C}_c$ quantifies the mismatch with the corresponding MC response template. This photon-count-only method provides a non-neural-network reference for evaluating the CNN performance. The SiPM positions therefore enter the template reconstruction implicitly through $r_c(\theta)$.

\subsection{CNN track reconstruction}

The primary track reconstruction method used for the final endpoint studies is a time-informed three-dimensional CNN trained on simulated detector-response samples~\cite{LeCun:1998Gradient,Aurisano:2016CNN,DUNE:2020kinematic}.
The network input is a voxelized detector volume containing event-dependent charge and timing-rank channels.
Three fixed coordinate channels encoding the SiPM $x$, $y$, and $z$ positions are concatenated with these event-dependent channels at the network input~\cite{Liu:2018CoordConv}.
This representation allows the CNN to learn detector-response patterns while retaining the spatial and temporal ordering of the observed light.

For measured waveforms, a charge-weighted mean time is calculated for each channel within a 480~ns prompt window. For simulated events, the corresponding channel quantity is the mean arrival time of the detected photons.
The channels are then ordered independently within each event from earliest to latest. The earliest channel is assigned rank 1, with increasing integer ranks assigned to later channels.
This timing-rank representation is invariant to an event-level common time offset and is therefore less sensitive to trigger jitter and absolute timing differences between simulation and data.
It also preserves the directional information carried by relative photon-arrival order,
which complements the charge pattern when distinguishing entry and exit regions.
The use of photon-arrival timing as topology information is also supported by scintillator-detector studies that exploit timing and detected light density to recover directional optical signatures~\cite{Caravaca:2017CHESS}.

\begin{figure}[htbp]
	\begin{center}
		\includegraphics[width=0.30\textwidth]{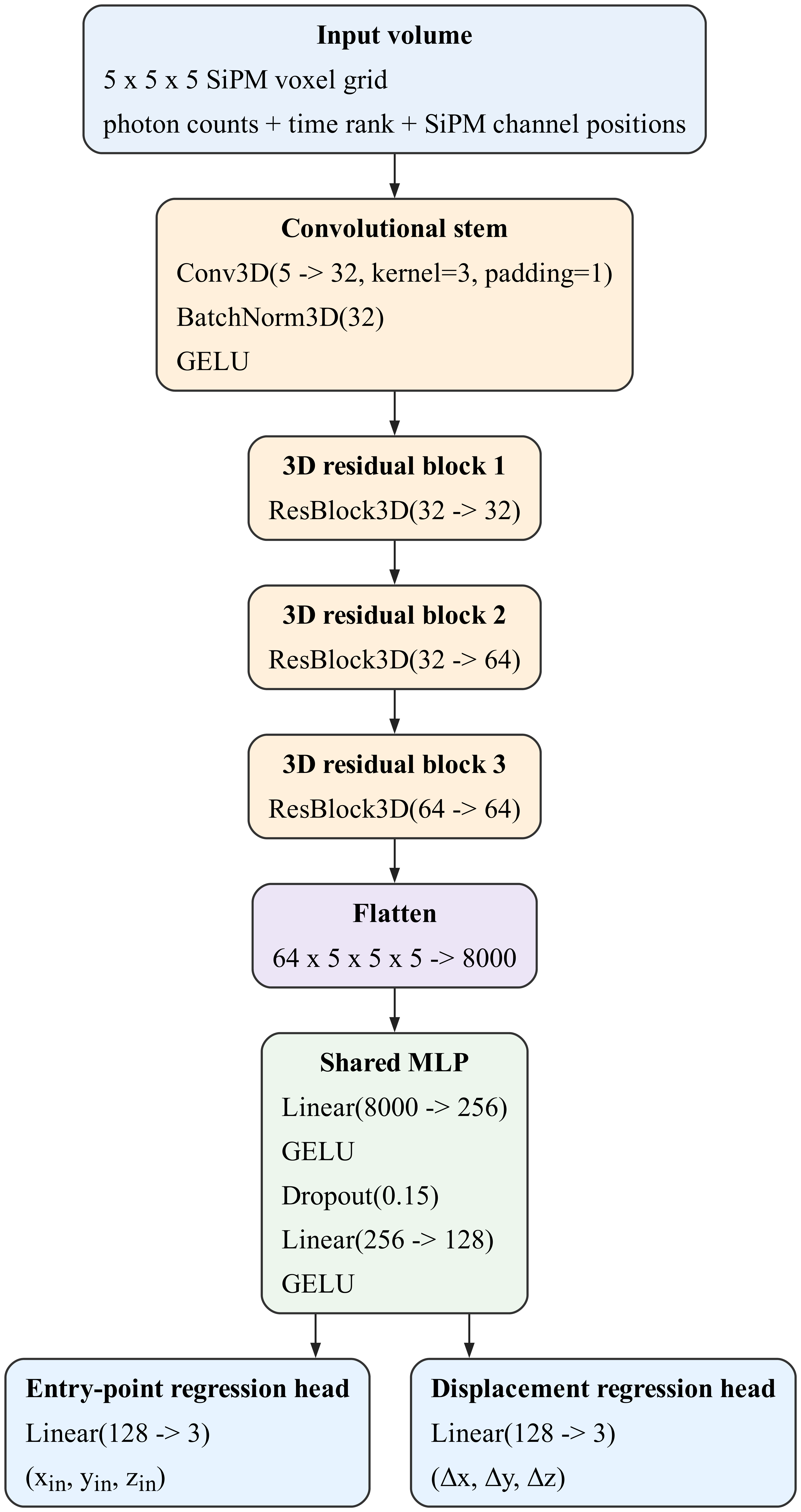}
	\end{center}
	\caption{
		Compact architecture of the time-informed CNN used for trajectory reconstruction.
		Three-dimensional convolutional and residual blocks extract local detector-response features,
		which are passed to dense regression heads for the entry point and displacement vector.
	}
	\label{fig:cnn}
\end{figure}

Figure~\ref{fig:cnn} shows the network architecture used for trajectory reconstruction.
The architecture comprises a three-dimensional convolutional stem, residual 3D convolutional blocks~\cite{He:2016ResNet},
	flattening and dense layers, and two regression heads.
Three-dimensional convolutions are used because neighboring SiPM channels carry correlated charge and timing information,
while the SiPM-position channels provide the physical sensor locations in the sparse lattice.
Residual connections facilitate optimization by improving gradient propagation and allowing each block to learn corrections to its input.
The output heads predict the particle entry point, $\mathbf r_{\rm entry}$, and the entry-to-exit displacement vector,

\begin{equation}
		\Delta\mathbf r=(\Delta x,\Delta y,\Delta z)
		=\mathbf r_{\rm exit}-\mathbf r_{\rm entry}.
\end{equation}

The reconstructed exit point is computed as

\begin{equation}
		\mathbf r_{\rm exit}^{\rm reco}
		=\mathbf r_{\rm entry}^{\rm reco}+\Delta\mathbf r^{\rm reco},
\end{equation}

with the final endpoint constrained to remain within the detector volume.

The CNN training dataset comprised through-going muon events selected from CRY and Geant4 simulations.
Events were retained only when complete truth information was available for both entry and exit coordinates and the total simulated photon count exceeded 15 photons.
This selection retained 152,551 events.
The sample was divided into training, validation, and test subsets with a 70:15:15 ratio,
as summarized in Table~\ref{tab:cnndataset}.
Input and target normalizations were fitted using only the training subset.
The model was trained with Huber loss~\cite{Huber:1964Robust} to reduce sensitivity to rare endpoint outliers and with stochastic gradient descent (SGD) using cosine annealing~\cite{Loshchilov:2017SGDR} to provide a smooth learning-rate schedule.

\begin{table}[htbp]
	\caption{
		CNN training-sample summary for the time-informed model used in the endpoint reconstruction study.
	}
	\label{tab:cnndataset}
	\centering
	\smallskip
	\begin{tabular}{lc}
			Quantity               & Value    \\
			\hline
			Selected simulated events & 152,551  \\
			Training events        & 106,785  \\
			Validation events      & 22,882   \\
			Test events            & 22,884   \\
			Split ratio            & 0.70 : 0.15 : 0.15 \\
			Optimization           & SGD with cosine annealing \\
			Loss function          & Huber            \\
		\end{tabular}
\end{table}

\subsection{Endpoint and trigger-region metrics}

Performance in simulation is quantified using the spatial distance between true and reconstructed endpoints and the angular difference between the true and reconstructed track directions.
For through-going muons, the entry- and exit-point errors, $d_{\rm entry}$ and $d_{\rm exit}$, are evaluated separately as

\begin{equation}
		d_{\mathrm{entry,exit}} =
	\left|\mathbf{r}_{\mathrm{reco}}-\mathbf{r}_{\mathrm{true}}\right|,
\end{equation}
where $\mathbf{r}_{\mathrm{reco}}$ and $\mathbf{r}_{\mathrm{true}}$ denote the reconstructed and simulated endpoint positions, respectively.

The angular error is defined as

\begin{equation}
	\alpha = \cos^{-1}\!\left(\hat{\mathbf u}_{\rm reco}\cdot\hat{\mathbf u}_{\rm true}\right),
	\qquad
	\hat{\mathbf u}=\frac{\mathbf r_{\rm exit}-\mathbf r_{\rm entry}}
	{\left|\mathbf r_{\rm exit}-\mathbf r_{\rm entry}\right|}.
\end{equation}

The external-trigger data do not provide event-by-event truth endpoint positions.
Instead, the reconstructed entry and exit points are compared with the known geometric acceptance of the external trigger counters.
The trigger-region consistency distance is defined as zero when the reconstructed point lies inside the rectangular counter acceptance. Otherwise, it is the shortest distance in the $x$-$y$ plane from the point to the nearest counter boundary. This one-sided metric measures only the distance beyond the finite trigger region and is insensitive to the spread of reconstructed points within that region.

\section{Reconstruction performance in simulation}

\subsection{Through-going muon endpoint resolution}

\begin{figure}[htbp]
	\begin{center}
		\includegraphics[width=0.8\textwidth]{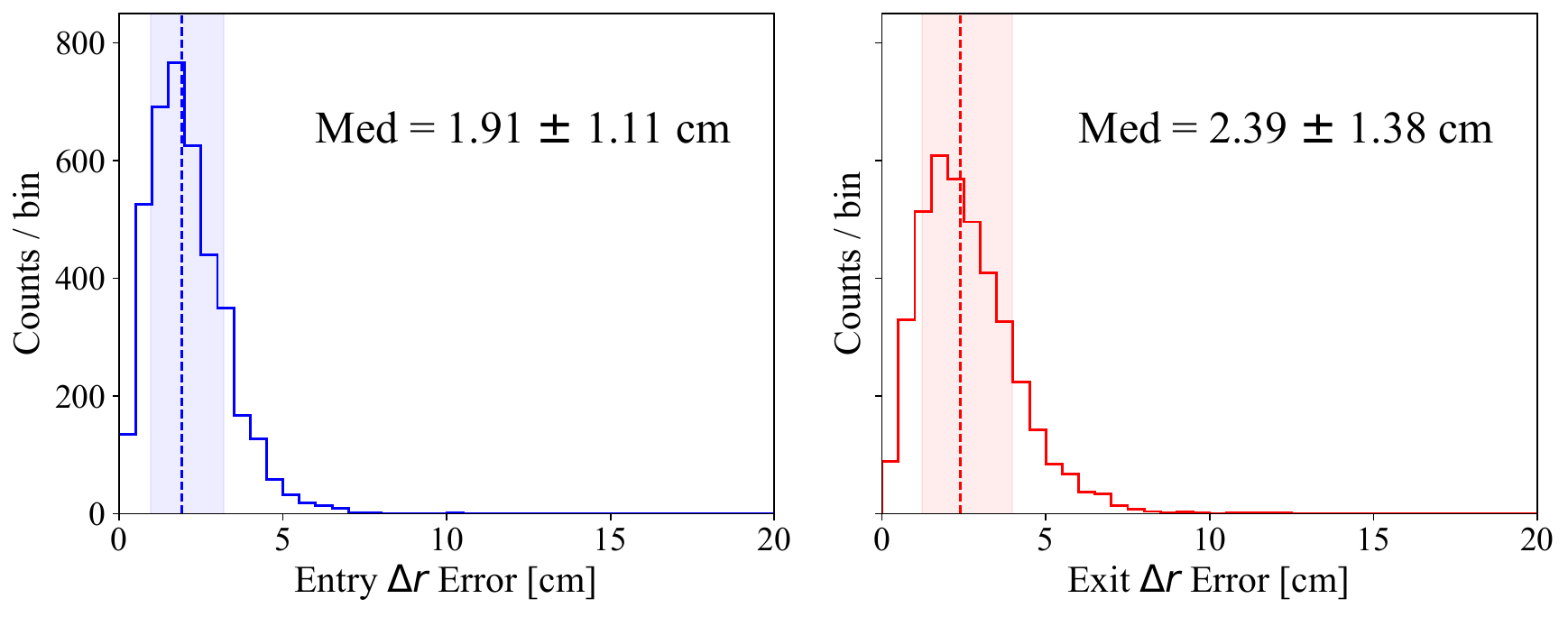}
	\end{center}
	\caption{
		Endpoint reconstruction residuals for simulated top-to-bottom through-going muons.
		The blue and red histograms show the three-dimensional entry- and exit-point residuals, respectively.
		Dashed lines indicate the median values and the shaded bands show the central 68.3\% intervals.
	}
	\label{fig:entryexit}
\end{figure}

Figure~\ref{fig:entryexit} shows the performance on a simulated test subset that was not used for model training or early stopping.
For these events, the true entry point lies on the top face and the true exit point lies on the bottom face of the detector.
This selection excludes side-entering and side-exiting events that are only partially constrained by the detector volume.
The plotted sample contains 3,967 selected events from the test subset.
The resulting median residuals are 1.91~cm for the entry point and 2.39~cm for the exit point, with half-widths of 1.11~cm and 1.38~cm for the central 68.3\% intervals.

\begin{figure}[htbp]
	\begin{center}
		\includegraphics[width=0.75\textwidth]{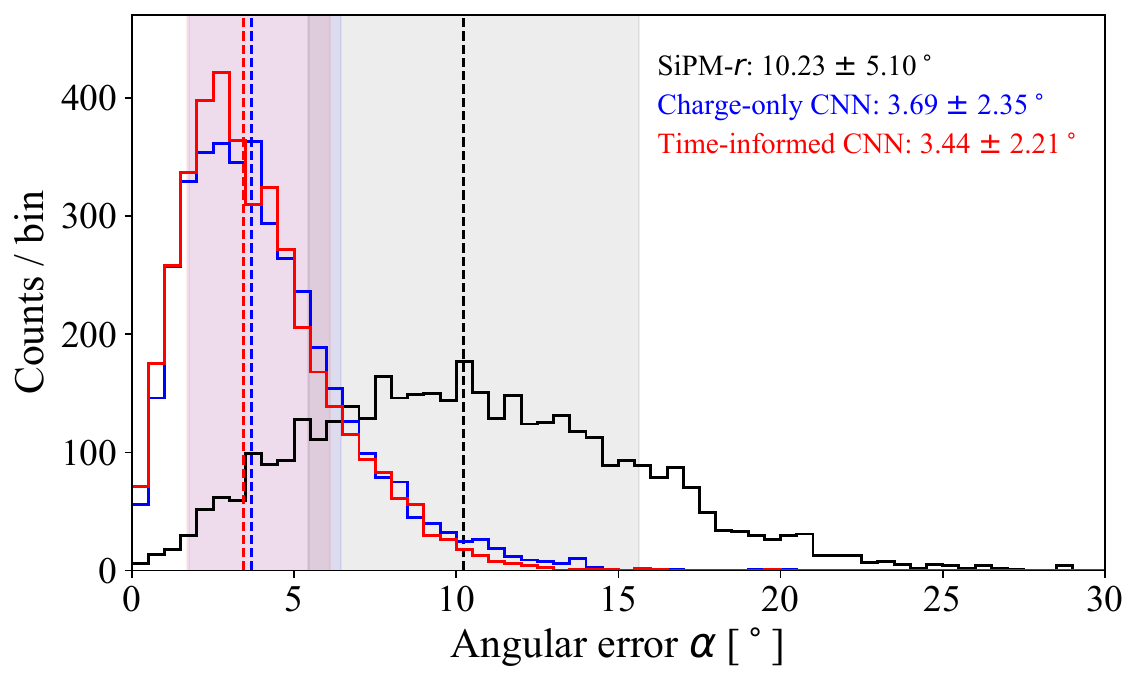}
	\end{center}
	\caption{
		Angular reconstruction errors for the SiPM-$r$ template, charge-only CNN, and time-informed CNN on the same 3,967 simulated top-to-bottom through-going muons.
			The black, blue, and red histograms show the SiPM-$r$ template, charge-only CNN, and time-informed CNN, respectively.
		Dashed lines indicate the median values and the shaded bands show the central 68.3\% intervals.
	}
	\label{fig:angularcomparison}
\end{figure}

The pair of reconstructed endpoints determines both the track position and direction. Figure~\ref{fig:angularcomparison} shows the corresponding angular-error distributions for the three reconstruction methods. For the time-informed CNN, the median angular error for the selected sample is $3.44^{\circ}$.

In the broader simulated through-going-muon sample before the top-to-bottom requirement,
the largest residuals are associated with corner-clipping tracks, low photon-count events,
	and events in which only a short portion of the charged-particle path is contained in the detector volume.

The larger exit residual is consistent with the output parameterization because uncertainties in both the reconstructed entry point and displacement vector, including their covariance, contribute to the exit-point error.

Table~\ref{tab:recomethodcomparison} compares the SiPM-$r$ template, a charge-only CNN, and the time-informed CNN on the same selected sample.
This three-way comparison separates the effects of the reconstruction model and timing information.

\begin{table}[htbp]
	\caption{
			Reconstruction performance for the SiPM-$r$ template, charge-only CNN, and time-informed CNN on the simulated test subset.
			All methods are evaluated on the same 3,967 top-to-bottom test events.
		Values are reported as the median $\pm$ the half-width of the central 68.3\% interval.
	}
	\label{tab:recomethodcomparison}
	\footnotesize
	\centering
	\smallskip
		\begin{tabular}{lccc}
				Method & Entry [cm] & Exit [cm] & Angular error [$^{\circ}$] \\
				\hline
				SiPM-$r$ template & $5.13 \pm 3.59$ & $7.58 \pm 3.86$ & $10.23 \pm 5.10$ \\
				Charge-only CNN & $2.01 \pm 1.17$ & $2.58 \pm 1.55$ & $3.69 \pm 2.35$ \\
				Time-informed CNN & $1.91 \pm 1.11$ & $2.39 \pm 1.38$ & $3.44 \pm 2.21$ \\
			\end{tabular}
\end{table}

The performance differences reflect both the reconstruction model and the information supplied to it.
The SiPM-$r$ template reconstruction treats each channel primarily through an independent photon-count-to-distance response,
so it cannot fully exploit local correlations among neighboring sensors or event-level timing patterns.
The charge-only CNN uses the same event-level photon counts while learning nonlinear spatial correlations across the fixed SiPM positions. Adding the timing-rank channel further reduces the median entry, exit, and angular errors by 5.3\%, 7.5\%, and 6.6\%, respectively.

\subsection{Starting-track vertex reconstruction}

For the positron starting-track sample, positrons were generated at positions sampled uniformly throughout the detector volume. Their initial directions were isotropic, and their kinetic energies were sampled uniformly between 0 and 50~MeV. The reconstruction target was the positron production vertex. Compared with through-going muons, these events produce shorter and more isotropically oriented charged-particle tracks. Their vertex reconstruction is therefore constrained mainly by the local photon-count pattern rather than by a long track crossing multiple detector planes.

The compact detector volume provides a demanding test of starting-track vertex reconstruction because 28.2\% of the generated positron tracks partially exit the detector volume.

\begin{figure}[htbp]
	\begin{center}
		\includegraphics[width=0.60\textwidth]{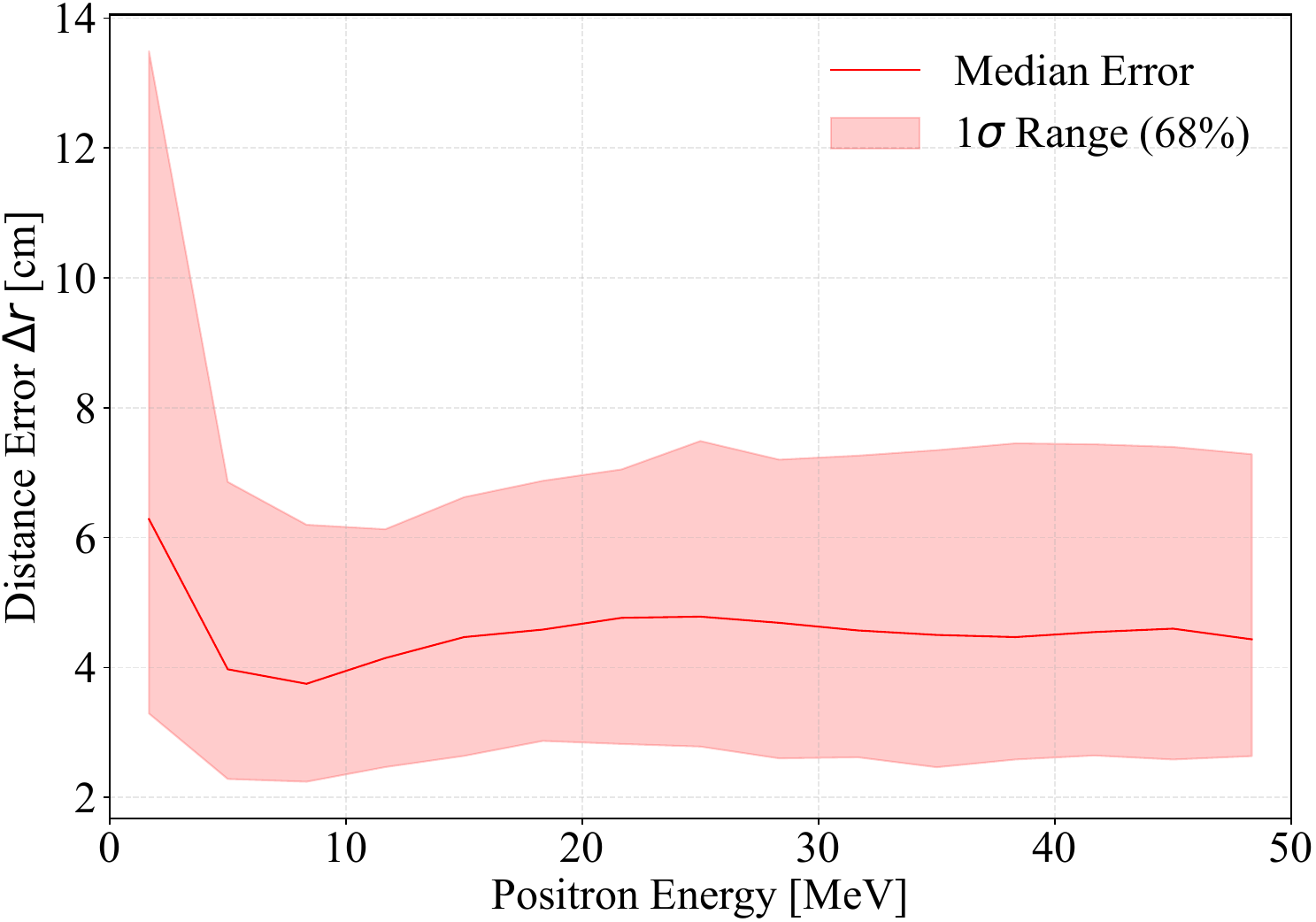}
	\end{center}
	\caption{
				Production-vertex residual for simulated positron starting-track events as a function of initial kinetic energy. The line shows the median residual, and the shaded band indicates the central 68.3\% interval.
	}
	\label{fig:pointreco}
\end{figure}

Figure~\ref{fig:pointreco} shows the production-vertex residual for the simulated positron starting-track events as a function of initial kinetic energy.

Across the studied energy range, the median production-vertex residual is about 4.5~cm.

\section{External-trigger validation with real data}

\subsection{External trigger setup}

\begin{figure}[htbp]
	\begin{center}
		\includegraphics[width=0.40\textwidth]{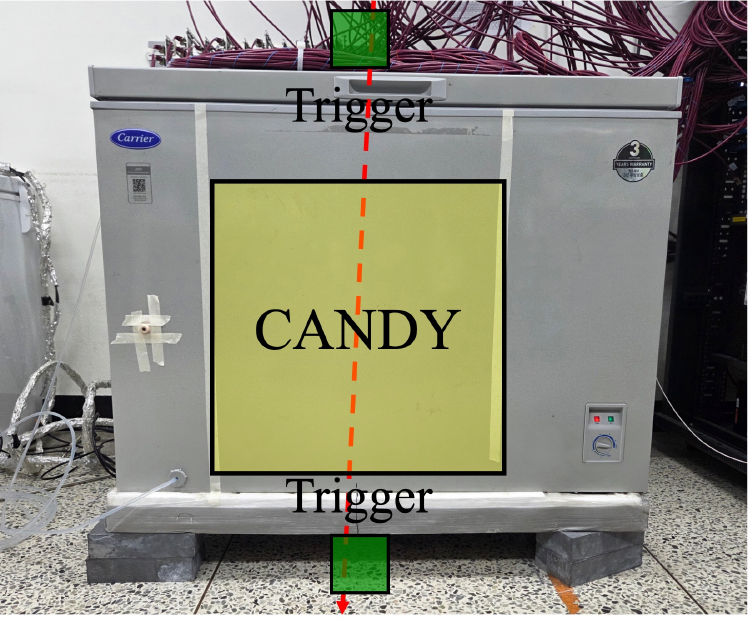}
	\end{center}
	\caption{
		External-trigger validation geometry and representative reconstructed through-going muon event.  The external scintillator counters define a narrow near-vertical trigger region, providing an independent geometric constraint on the expected muon trajectory.
	}
	\label{fig:verticalsetup}
\end{figure}

To experimentally validate directional reconstruction, two external scintillator trigger counters were positioned above and below the detector to select near-vertical through-going cosmic-ray muons. The trigger counters each had an active area of about 25~cm$^2$ and were separated vertically by 1~m. During normal operation, CANDY passively records cosmic-ray events that satisfy its internal multiplicity trigger. Coincidences with the external counters define a validation sample with an additional geometric constraint. Figure~\ref{fig:verticalsetup} shows the trigger geometry.

\subsection{Event selection and reconstruction}

For the real-data reconstruction sample, events were required to satisfy the external trigger condition and $3000<N_{\mathrm{PE,prompt}}^{\mathrm{tot}}<30000$. Here, $N_{\mathrm{PE,prompt}}^{\mathrm{tot}}$ is obtained by converting the prompt ADC integral in each channel to photoelectrons using the channel-dependent gain calibration and then summing over all channels.

The data stream also included periodic monitoring triggers issued every 30 minutes to track and correct long-term baseline variations. Consequently, the total event counts in Table~\ref{tab:realtriggersummary} include these monitoring triggers and should not be interpreted as the number of through-going muon candidates. The selected events were reconstructed with the same time-informed CNN.

\begin{figure}[htbp]
	\begin{center}
		\includegraphics[width=0.70\textwidth]{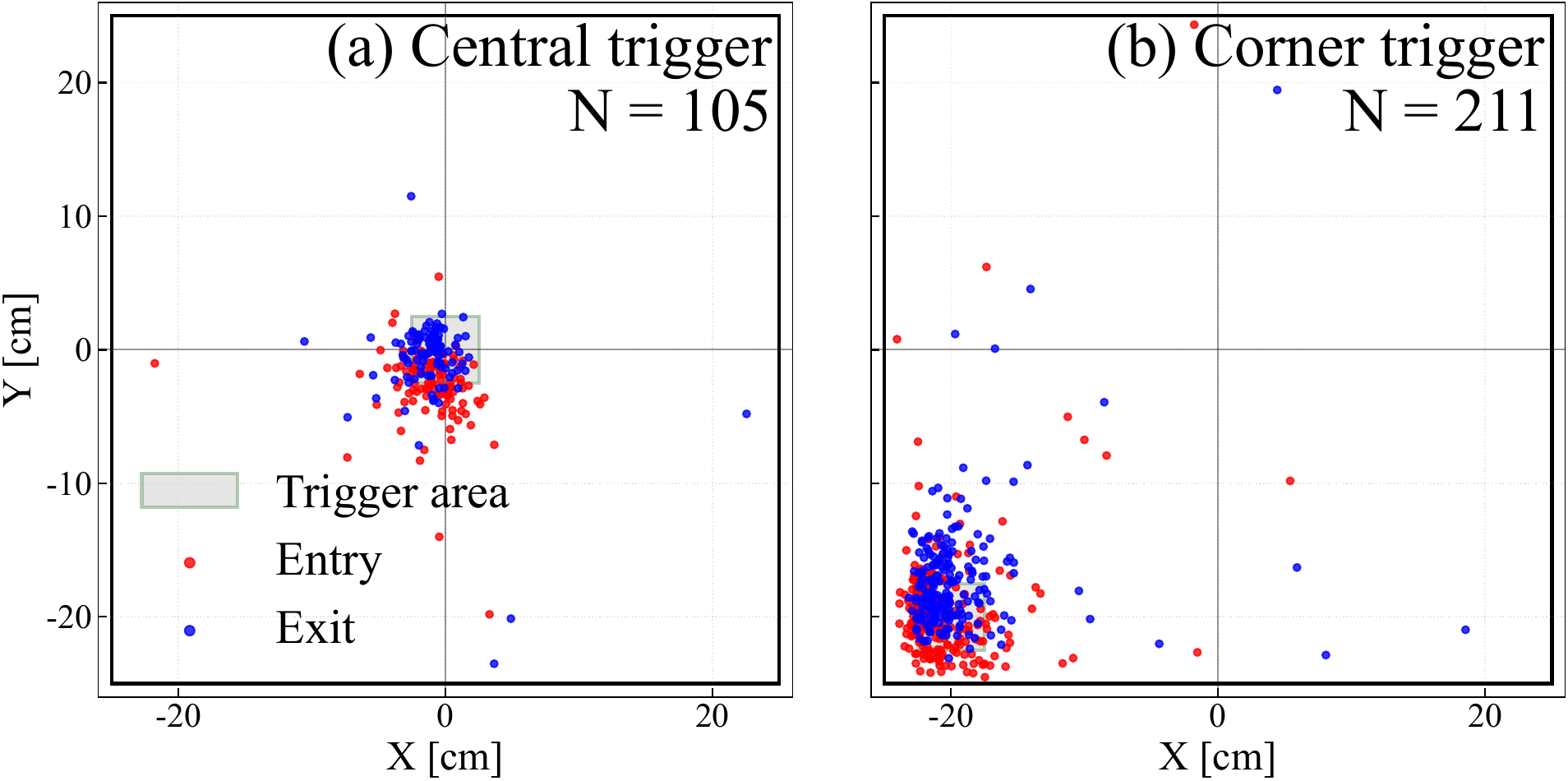}
	\end{center}
	\caption{
				Reconstructed entry and exit positions for externally triggered real-data events after the trigger and prompt-charge selections.
		The left panel shows the central-trigger dataset with the trigger centered at $(0,0)$~cm, and the right panel shows the corner-trigger dataset with the trigger centered at $(-20,-20)$~cm.
		The selected samples contain 105 events for the central-trigger data and 211 events for the corner-trigger data.
		Red and blue points indicate the CNN-reconstructed entry and exit positions, respectively.
	}
	\label{fig:realtriggerreco}
\end{figure}

\begin{table}[htbp]
	\caption{
			Data-taking, event-selection, and reconstruction summary for the real-data samples shown in Fig.~\ref{fig:realtriggerreco}. Total event counts include periodic monitoring triggers. The entry and exit values are coordinate-wise medians of the reconstructed $x$-$y$ positions.
	}
	\label{tab:realtriggersummary}
	\footnotesize
	\centering
	\smallskip
	\begin{tabular}{lcc}
			Quantity & Central trigger & Corner trigger \\
			\hline
			Live time [s] & 262,027 & 466,112 \\
			Total events & 490 & 1000 \\
			Reconstructed events & 105 & 211 \\
			Entry median $(x,y)$ [cm] & $(-0.83,-2.49)$ & $(-21.21,-20.07)$ \\
			Exit median $(x,y)$ [cm] & $(-1.01,-0.11)$ & $(-20.68,-18.27)$ \\
		\end{tabular}
\end{table}

\begin{figure}[htbp]
	\begin{center}
		\includegraphics[width=0.9\textwidth]{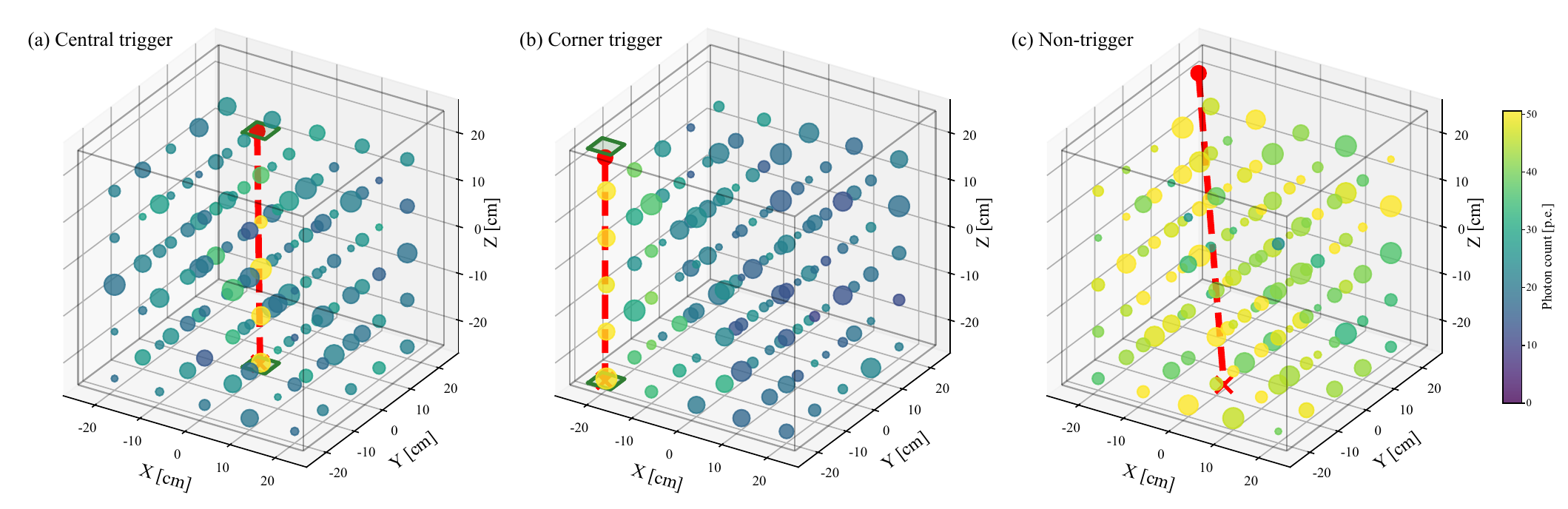}
	\end{center}
	\caption{
				Representative reconstructed real-data event displays for (a) the central-trigger sample, (b) the corner-trigger sample, and (c) a cosmic-ray muon candidate selected without external trigger counters. The SiPM marker color encodes the photon count, with brighter colors corresponding to larger photon counts. A common photon-count color scale is used for all three panels. The marker size encodes the charge-weighted mean hit time after event-local normalization, with larger markers corresponding to earlier channels. The red dashed line indicates the CNN-predicted trajectory, and the green squares in panels (a) and (b) show the external trigger-counter positions.
	}
	\label{fig:verticalreco}
\end{figure}

The reconstructed coordinate medians in Table~\ref{tab:realtriggersummary} lie within or near the corresponding trigger-defined regions. Representative event displays for the two trigger configurations and a muon candidate selected without external trigger counters are shown in Fig.~\ref{fig:verticalreco}.

As shown in Fig.~\ref{fig:realtriggerreco}, the reconstructed endpoint distributions overlap the corresponding trigger-defined regions but retain a finite spread and visible tails. For the combined central- and corner-trigger samples, the median entry- and exit-point region distances are 0.29~cm and 0.00~cm, respectively. Because every point inside a finite trigger region is assigned a distance of zero, the zero median exit distance means only that at least half of the reconstructed exit points lie inside their corresponding regions. These values therefore quantify acceptance consistency rather than sub-centimeter endpoint resolution or perfect event-by-event reconstruction.

\section{Discussion and outlook}

The reconstruction performance is driven by the spatially distributed photon-count pattern and event-local timing order measured across the in-liquid SiPM array.  Unlike a boundary-readout detector, the in-liquid sensor lattice samples the scintillation field before the optical information is fully averaged by long transport paths.  Consequently, even a sparse lattice can retain topology information when the reconstruction uses both the three-dimensional charge pattern and the relative time ordering of hit channels.

The charge-only CNN already outperforms the SiPM-$r$ template, showing that much of the performance gain is obtained without timing information.
Adding timing rank reduces the median reconstruction errors by a further 5--8\%.
This improvement shows that relative arrival order provides useful information beyond the photon-count pattern.

The validation strategy separates detector-response modeling, reconstruction performance on simulated events with truth information, and real-data trigger consistency.
The photon-count data-to-MC comparison shows that the dominant light-response scale is reproduced after an effective scale correction,
while the simulation test evaluates endpoint reconstruction against truth labels.
The external-trigger sample does not provide event-by-event truth positions,
but it tests a different question: whether the simulation-trained model reconstructs measured events in a way that is geometrically consistent with independently placed trigger counters.
Agreement in this real-data test supports the use of the simulation-trained model beyond a purely simulated performance study.

The photon-count scale correction absorbs multiple detector-response uncertainties, including photon-detection efficiency,
optical-model parameters, scintillation yield, and calibration effects.
The current finite Monte Carlo exposure also limits the high-photon-count tail,
	where multiple-muon and accidental coincidences can contribute to the data. These limitations do not directly affect the quoted simulation-only endpoint metrics.
However, these aspects should be addressed for precision detector-response modeling and for extending the method to broader event classes.

The positron starting-track study extends the evaluation beyond through-going tracks to production-vertex reconstruction for shorter tracks in the compact detector volume.
The comparison of CRY-generated through-going and stopping muons in Fig.~\ref{fig:mcselection} also provides a useful reference for choosing stopping-muon-enriched samples,
while the detailed delayed Michel-electron waveform analysis is treated separately.

A larger detector may provide longer track lengths, but greater sensor spacing and longer optical paths can reduce the available local information.
The dependence of reconstruction performance on detector size and sensor spacing must therefore be evaluated before extrapolating the present results.

\section{Conclusion}

A sparse in-liquid SiPM lattice provides sufficient spatial and timing information for charged-particle reconstruction in a homogeneous liquid scintillator detector.
On the simulated top-to-bottom muon test sample, the time-informed CNN achieves median entry and exit residuals of 1.91~cm and 2.39~cm and a median angular error of $3.44^{\circ}$.
It outperforms the SiPM-$r$ template on the same events, and the charge-only comparison shows an additional benefit from timing rank.

For externally triggered cosmic-ray muon data, the reconstructed endpoints are consistent with the trigger-defined regions.
Simulated positron starting-track events also yield a median production-vertex residual of about 4.5~cm.
These results support the feasibility of reconstructing both through-going tracks and localized starting vertices with sparse in-liquid readout.
Extending this approach to larger homogeneous detectors will require further studies of optical response, timing calibration, and sensor spacing~\cite{Askins:2020Theia}.

\acknowledgments
This research was supported by the Chung-Ang University Research Scholarship Grants in 2025
and by the National Research Foundation of Korea (NRF) grant funded
by the Korean government (MSIT) (RS-2021-NR058750 and RS-2024-00438814).

\bibliographystyle{JHEP}
\bibliography{biblio}

@article{CANDYLS,
    author = "Kwak, M. S. and Chung, J. S. and Ha, C. and Huang, T. Z. and Kim, J. Y. and Kim, S. A. and Kimku, H. and Koh, B. C. and Lee, H. S. and Lee, S. and Lee, Y. J. and Seo, J.",
    title = "{Characterization of immersed SiPM arrays in liquid scintillator between room temperature and $-30\,^{\circ}\mathrm{C}$}",
    eprint = "2609.25534",
    archivePrefix = "arXiv",
    primaryClass = "physics.ins-det",
    year = "2026"
}

@inproceedings{KamLAND:2004overview,
    author = "Suekane, F. and Iwamoto, T. and Ogawa, H. and Tajima, O. and Watanabe, H.",
    title = "{An Overview of the KamLAND 1-kiloton Liquid Scintillator}",
    booktitle = "{Neutrino Oscillations and Their Origin}",
    eprint = "physics/0404071",
    archivePrefix = "arXiv",
    doi = "10.1142/9789812703101_0018",
    year = "2004"
}

@article{Borexino:2008detector,
    author = "Alimonti, G. and others",
    collaboration = "Borexino",
    title = "{The Borexino detector at the Laboratori Nazionali del Gran Sasso}",
    eprint = "0806.2400",
    archivePrefix = "arXiv",
    primaryClass = "physics.ins-det",
    doi = "10.1016/j.nima.2008.11.076",
    journal = "Nucl. Instrum. Meth. A",
    volume = "600",
    pages = "568--593",
    year = "2009"
}

@article{SNOplus:2021experiment,
    author = "Albanese, V. and others",
    collaboration = "SNO+",
    title = "{The SNO+ Experiment}",
    eprint = "2104.11687",
    archivePrefix = "arXiv",
    primaryClass = "physics.ins-det",
    doi = "10.1088/1748-0221/16/08/P08059",
    journal = "JINST",
    volume = "16",
    number = "08",
    pages = "P08059",
    year = "2021"
}

@article{NEOS:2016,
    author = "Ko, Y. J. and others",
    collaboration = "NEOS",
    title = "{Sterile Neutrino Search at the NEOS Experiment}",
    eprint = "1610.05134",
    archivePrefix = "arXiv",
    primaryClass = "hep-ex",
    doi = "10.1103/PhysRevLett.118.121802",
    journal = "Phys. Rev. Lett.",
    volume = "118",
    number = "12",
    pages = "121802",
    year = "2017"
}

@article{DUNE:2020kinematic,
    author = "Liu, Junze and Ott, Jordan and Collado, Julian and Jargowsky, Benjamin and Wu, Wenjie and Bian, Jianming and Baldi, Pierre",
    title = "{Deep-Learning-Based Kinematic Reconstruction for DUNE}",
    eprint = "2012.06181",
    archivePrefix = "arXiv",
    primaryClass = "physics.ins-det",
    year = "2020"
}

@article{JUNO:2021vlw,
    author = "Abusleme, A. and others",
    collaboration = "JUNO",
    title = "{JUNO physics and detector}",
    eprint = "2104.02565",
    archivePrefix = "arXiv",
    primaryClass = "hep-ex",
    doi = "10.1016/j.ppnp.2021.103927",
    journal = "Prog. Part. Nucl. Phys.",
    volume = "123",
    pages = "103927",
    year = "2022"
}

@article{LiquidO:2021,
    author = "Cabrera, A. and others",
    title = "{Neutrino Physics with an Opaque Detector}",
    eprint = "1908.02859",
    archivePrefix = "arXiv",
    primaryClass = "physics.ins-det",
    doi = "10.1038/s42005-021-00763-5",
    journal = "Commun. Phys.",
    volume = "4",
    pages = "273",
    year = "2021"
}

@article{Askins:2020Theia,
    author = "Askins, M. and others",
    title = "{THEIA: an advanced optical neutrino detector}",
    eprint = "1911.03501",
    archivePrefix = "arXiv",
    primaryClass = "physics.ins-det",
    doi = "10.1140/epjc/s10052-020-7977-8",
    journal = "Eur. Phys. J. C",
    volume = "80",
    number = "5",
    pages = "416",
    year = "2020"
}

@article{Caravaca:2017CHESS,
    author = "Caravaca, J. and Descamps, F. B. and Land, B. J. and Orebi Gann, G. D. and Wallig, J. and Yeh, M.",
    title = "{An experiment to demonstrate separation of Cherenkov and scintillation signals}",
    eprint = "1610.02029",
    archivePrefix = "arXiv",
    primaryClass = "physics.ins-det",
    doi = "10.1140/epjc/s10052-017-4759-x",
    journal = "Eur. Phys. J. C",
    volume = "77",
    number = "12",
    pages = "811",
    year = "2017"
}

@article{GEANT4:2002zbu,
    author = "Agostinelli, S. and others",
    collaboration = "GEANT4",
    title = "{GEANT4 - A Simulation Toolkit}",
    reportNumber = "SLAC-PUB-9350, FERMILAB-PUB-03-339, CERN-IT-2002-003",
    doi = "10.1016/S0168-9002(03)01368-8",
    journal = "Nucl. Instrum. Meth. A",
    volume = "506",
    pages = "250--303",
    year = "2003"
}

@article{GEANT4:2016lxb,
    author = "Allison, J. and others",
    title = "{Recent developments in Geant4}",
    doi = "10.1016/j.nima.2016.06.125",
    journal = "Nucl. Instrum. Meth. A",
    volume = "835",
    pages = "186--225",
    year = "2016"
}

@inproceedings{Hagmann:2007CRY,
    author = "Hagmann, Chris and Lange, David and Wright, Douglas",
    title = "{Cosmic-ray shower generator (CRY) for Monte Carlo transport codes}",
    booktitle = "{2007 IEEE Nuclear Science Symposium Conference Record}",
    volume = "2",
    pages = "1143--1146",
    doi = "10.1109/NSSMIC.2007.4437209",
    year = "2007"
}

@article{Renker:2006SiPM,
    author = "Renker, Dieter",
    title = "{Geiger-mode avalanche photodiodes, history, properties and problems}",
    doi = "10.1016/j.nima.2006.05.060",
    journal = "Nucl. Instrum. Meth. A",
    volume = "567",
    pages = "48--56",
    year = "2006"
}

@article{LeCun:1998Gradient,
    author = "LeCun, Yann and Bottou, L{\'e}on and Bengio, Yoshua and Haffner, Patrick",
    title = "{Gradient-based learning applied to document recognition}",
    doi = "10.1109/5.726791",
    journal = "Proc. IEEE",
    volume = "86",
    number = "11",
    pages = "2278--2324",
    year = "1998"
}

@article{Aurisano:2016CNN,
    author = "Aurisano, A. and Radovic, A. and Rocco, D. and Himmel, A. and Messier, M. D. and Niner, E. and Pawloski, G. and Psihas, F. and Sousa, A. and Vahle, P.",
    title = "{A Convolutional Neural Network Neutrino Event Classifier}",
    eprint = "1604.01444",
    archivePrefix = "arXiv",
    primaryClass = "hep-ex",
    doi = "10.1088/1748-0221/11/09/P09001",
    journal = "JINST",
    volume = "11",
    number = "09",
    pages = "P09001",
    year = "2016"
}

@inproceedings{He:2016ResNet,
    author = "He, Kaiming and Zhang, Xiangyu and Ren, Shaoqing and Sun, Jian",
    title = "{Deep Residual Learning for Image Recognition}",
    booktitle = "{2016 IEEE Conference on Computer Vision and Pattern Recognition (CVPR)}",
    eprint = "1512.03385",
    archivePrefix = "arXiv",
    primaryClass = "cs.CV",
    doi = "10.1109/CVPR.2016.90",
    pages = "770--778",
    year = "2016"
}

@inproceedings{Liu:2018CoordConv,
    author = "Liu, Rosanne and Lehman, Joel and Molino, Piero and Petroski Such, Felipe and Frank, Eric and Sergeev, Alex and Yosinski, Jason",
    title = "{An Intriguing Failing of Convolutional Neural Networks and the CoordConv Solution}",
    booktitle = "{Advances in Neural Information Processing Systems}",
    eprint = "1807.03247",
    archivePrefix = "arXiv",
    primaryClass = "cs.CV",
    year = "2018"
}

@article{Huber:1964Robust,
    author = "Huber, Peter J.",
    title = "{Robust Estimation of a Location Parameter}",
    doi = "10.1214/aoms/1177703732",
    journal = "Annals of Mathematical Statistics",
    volume = "35",
    number = "1",
    pages = "73--101",
    year = "1964"
}

@inproceedings{Loshchilov:2017SGDR,
    author = "Loshchilov, Ilya and Hutter, Frank",
    title = "{SGDR: Stochastic Gradient Descent with Warm Restarts}",
    booktitle = "{International Conference on Learning Representations}",
    eprint = "1608.03983",
    archivePrefix = "arXiv",
    primaryClass = "cs.LG",
    year = "2017"
}

@book{knoll,
  title={Radiation detection and measurement},
  author={Knoll, Glenn F.},
  year={2010},
  publisher={John Wiley and Sons, New York}
}

@misc{sipm,
    author = "{Hamamatsu Photonics K.K.}",
    title = "{MPPC S13360 series data sheet}",
    year = "2024",
    url = "https://www.hamamatsu.com/eu/en/product/optical-sensors/mppc"
}

@article{tyvek,
  author={Janecek, Martin},
  journal={IEEE Transactions on Nuclear Science}, 
  title={Reflectivity Spectra for Commonly Used Reflectors}, 
  year={2012},
  volume={59},
  number={3},
  pages={490-497},
  doi={10.1109/TNS.2012.2183385}
  }

\end{document}